\documentstyle[prl,aps,floats,graphics,twocolumn,epsf]{revtex}
\begin{document}
\def\be{\begin{equation}}
\def\ee{\end{equation}}
\def\bear{\begin{eqnarray}}
\def\eear{\end{eqnarray}}
\def\E{{\rm e}}
\def\bearst{\begin{eqnarray*}}
\def\eearst{\end{eqnarray*}}
\def\peleven{\parbox{11cm}}
\def\peffec{\peight{\bearst\eearst}\hfill\peleven}
\def\pspace{\peight{\bearst\eearst}\hfill}
\def\ptwelve{\parbox{12cm}}
\def\peight{\parbox{8mm}}
\twocolumn[\hsize\textwidth\columnwidth\hsize\csname @twocolumnfalse\endcsname
\title
{Scale-invariant Perturbation of the Friedmann Universe}
\author
{Elcio Abdalla$^a$ and 
Roya Mohayaee$^{ab}$}
\address
{$^a$\it Instituto de 
F\'\i sica-USP, S\~ao Paulo\\
$^b$ HEP, ICTP, Trieste \\
}
\date{9/10/97}
\maketitle

%%%%%%%%%%%%%%%%%%%%%%%%%%%%%%%%%%%%%%%%%%%%%%%%%%%%%%
%ABSTRACT
%%%%%%%%%%%%%%%%%%%%%%%%%%%%%%%%%%%%%%%%%%%%%%%%%%%%%%
\begin{abstract}

We use de Vaucouleurs' power-law density-distance relation, 
to study a hierarchical perturbation of the Friedmann universe. 
We solve the Einstein equation
and obtain the density contrast and the amplification factor
for the perturbation. It is shown that,  
scale-invariant inhomogeneities decay in Einstein-de-Sitter
universe.
On the contrary, in an open 
universe, the inhomogeneities grow. For low values of $\Omega$, 
amplification peaks sharply and the fluctuations can grow
by up to a factor of $10^{13}$ from the 
recombination to the present time. 
Our analysis of the closed universe further confirms
that, unlike the common belief that
perturbations grow faster with increasing $\Omega$,
scale-invariant perturbations amplify
with decreasing $\Omega$.
This result is consistent with the very low 
density expected for a hierarchical universe. 

\end{abstract}
\hspace{.2in}

]
It is unanimously agreed that, up to a 
certain scale, the distribution
of matter in the universe is inhomogeneous 
\cite{peeble}. 
However, the scale at which the distribution 
shifts from a power law
to a random behaviour is a subject of strong
controversy \cite{pietronero,pietroneroreport,schramm,davis}.
For scales up to $100$ pc, it is well-established that the 
interstellar medium obeys a scaling 
law \cite{turbulence,devega}.
Further up the scale to $150$ Mpc, numerous red shift surveys
indicate the existence of large voids and filaments \cite{voids}.
It is argued that these structures have a 
fractal distribution \cite{pietronero,pietroneroreport}.
These facts have brought doubts about the hypothesis 
of the homogeneity, which is at the wellspring
of the standard cosmology \cite{weinberg,kolb,lucchin}.

It remains an ambitious task to study the dynamics of a fractal
universe. However, analytic inhomogeneous universes
and also, perturbatively, the evolution of inhomogeneities
in a homogeneous background
have been studied 
extensively.
De Vaucouleurs was among the first to study
a hierarchical distribution and the possibility of a universal
density-distance power law in cosmology \cite{vaucouleurs}.
The relativistic inhomogeneous cosmological
models and specially Tolman spacetime have also been
studied widely \cite{toleman,krasinski}. Tolman's dust 
solution has been used to 
model a relativistic hierarchical (fractal) 
cosmology compatible with the observational analysis of 
the redshift surveys \cite{ribeiro}.
In the perturbative approach,
the growth of 
small seeds of inhomogeneities is 
studied mainly in the 
context of the structure formation \cite{peeble,weinberg,kolb,lucchin}.
The large scale structure of the universe is 
believed to have grown, due to gravitational instability,
from small primordial density 
perturbations which are fully characterized 
by the density contrast.

In this article, we attempt to bring these ideas together.
A fractal structure is necessarily 
scale-invariant, eventhough the reverse statement is not true. 
We consider a scale-invariant spherically symmetric
inhomogeneous, but isotropic, universe which 
allows a non-vanishing pressure. 
The metric is the
Friedmann-Robertson-Walker metric with
radial-dependent scale factor 
and is contained in the Tolman solution for dust.
To solve the Einstein equation, we expand the scale
factor around the Friedmann scale 
factor by a small scale-invariant fluctuation.
The choice of a scale-invariant function for the density
perturbation is inspired by de Vaucouleurs'
original work on modeling a 
hierarchical universe \cite{vaucouleurs}.
We take his power-law density-distance relation to 
represent our perturbation
and aim at establishing if such a density 
fluctuation has any chance of growing 
into large scale structures we observe today.
Although the growth of a scale-invariant seed
does not imply a fully scale-invariant 
or a fractal universe, 
its decay would rule out such distributions 
as viable cosmological models.

To account for the inhomogeneity of space-time, we use the spherically
symmetric metric,
\bear
& &{\rm diag} \left(g_{\mu\nu}\right)=\nonumber\\
& &\left( -1,\, {{R_p}^2(t,r)\over 1-kr^2},\, {R_p}^2(t,r)r^2,\, 
{R_p}^2(t,r)r^2 \sin^2\theta\right),\label{metric} %(1)
\eear
where the scale factor $R_p(t,r)$ is a function of both
time and coordinate. The function $(1-kr^2)^{-1}$
multiplying ${R_p}^2(t,r)$ allows for the Friedman-Robertson-Walker
metric to be recovered when $R_p$ is space-coordinate independent.
We assume that the scale factor differs from 
the background homogeneous scale factor $R(t)$ by 
a scale-invariant perturbation term and write
\be
R_p(t,r)=R(t)+\delta R(t)r^{-\gamma}, \label{r}%(2)
\ee
where the comoving radial coordinate has been used. 
The value of the fractal 
codimension $\gamma$ has been recently proposed
to be about $1$ \cite{pietroneroreport} whereas 
it was originally believed to be approximately $2$ \cite{peeble}.
Similarly, we expand around the homogeneous background 
matter distribution $\rho$ with the scale-invariant
fluctuation to obtain the density function
\be
\rho_p(t,r)=\rho(t) +
\delta \rho(t) r^{-\gamma},\label{densityequation} %(4)
\ee
and the radial pressure 
\be
P_r(t,r)=\omega \rho(t) +\alpha \delta\rho(t)
r^{-\gamma}, \label{radialpressure} %(5)
\ee
where $\alpha$ is not 
necessarily 
equivalent to $\omega$. 
For practical purposes, however, it 
serves to consider an approximately 
adiabatic perturbation where $\alpha$ 
is almost equal to $\omega$.
In general, 
$\alpha =\omega =v_s^2$, where $v_s$ is 
the velocity of sound which
is assumed to be unaffected 
by the perturbation
\cite{weinberg,kolb,lucchin}.

The expression (\ref{r}) for the 
scale factor is chosen because it subsequently implies 
(\ref{densityequation}) for the density, 
which has the same form as that suggested by de Vaucouleurs
\cite{vaucouleurs}. 
The radial
dependence of the scale factor 
leads to the same
radial dependence for the density 
through Einstein equation. This 
scale-invariant behaviour is transfered to the 
pressure through the equation of state.
Furthermore, the scale factor and the density 
perturbations are required, by the energy-momentum 
conservation equation, to have the same fractal 
codimension.

The perturbation of the 
scale factor, $(\ref{r})$, is a genuine 
perturbation of the metric and not just a gauge 
mode. It is
not difficult to see that a solution $\xi_\mu$ of $\delta g_{\mu\nu}
=\xi_{\mu ,\nu} +\xi_{\nu ,\mu}$,
such that $\delta g_{0r}=0$ and
$\delta g_{rr}\sim t^ar^{-\gamma}$ 
for a general $a$, cannot 
be constructed \cite{ellis}.

Our perturbation is only valid for
comoving distances larger than the correlation length, where the
distribution is homogeneous and an average density is
defined. The correlation length has the controversial
minimum value of 5 Mpc and maximum value of 25 Mpc for 
some of the 3D galaxy catalogues
\cite{peeble,pietronero,pietroneroreport,davis,voids}.
For distances smaller than the correlation length, the 
fluctuation is large and
the density can be non-analytic with no specific average value
\cite{pietronero}. 
We are also assuming that crossover to homogeneity 
is reached at some stage. Otherwise, the notion of 
correlation length is meaningless and the present 
analytic approach is not applicable \cite{pietroneroreport}.

The Einstein equation is written for
the metric (\ref{metric}) and the perturbations
(\ref{r},\ref{densityequation},\ref{radialpressure}).
The zeroth-order
contributions, {\it i.e.}\  those with vanishing fractal codimension,
to the Einstein equation give the Friedmann equations.  
The first-order contributions lead to 
the second-order differential equation
\be
\ddot Y + \left(-6\pi G(\alpha+1)(\alpha\rho - P)-A_3 
{k\over R^2}\right) Y=0 \label{yequation} %(10)
\ee
where 
\bear
Y(R)&=& R^{-(3\alpha+1)/2} \delta R, \nonumber\\
A_3&=&(1+\gamma)\left[1-\alpha(\gamma-3)\right]-\frac 3 4
(\alpha+1)(3\alpha+1). %(11)
\eear

In the Einstein-de Sitter model, {\it i.e.} for $k =0$, 
the differential equation (\ref{yequation}) has exact solutions 
which simplify to
\be
\delta R\sim t^x\quad , \quad x=-{3\alpha +1\over 3(1+\omega)} +\frac 1 2 \pm
\frac 1 2 \left\vert {\omega -2\alpha -1\over \omega +1}\right\vert
\label{asympk=0}
\ee
at large times.
The scale factor fluctuation grows 
when $(\alpha +1)$ and $(\omega +1)$ have the same
sign and $2\vert \alpha +1\vert \ge \vert \omega +1\vert$.

The solutions (\ref{asympk=0}) can be substituted back 
in the time-time component of
the Einstein equation to find the 
perturbation in the density. The density contrast 
is subsequently given by
\be
\delta(t)={\delta\rho \over \rho}=
t^{-2\alpha+\omega-1\over 1+\omega}
\ee
which shows that our perturbation 
does not grow. Specifically, in matter or 
radiation -dominated eras the perturbation decays as
$t^{-1}$ which is the same as
the standard decaying mode \cite{weinberg,kolb,lucchin}.
Indeed, a simple calculation of the amplification 
factor, $\delta(t_0)/\delta(t)$,
 shows that 
the perturbation decays by a factor of $10^{-5}$
in the pressureless era 
({\it i.e.}, for $t/t_0\approx 10^5$).
The second solution for the density
contrast is just a constant.
The standard growing mode
of $t^{2/3}$ in the matter-dominated era
and $t$ in the radiation-dominated era \cite{weinberg,kolb,lucchin} 
do not arise in our scheme.
In order to recover these
results, we
need go to higher perturbative orders and
expand the scale factor (and subsequently density and pressure) 
in series of inverse powers of $r$, {\it i.e.} as,
\be
R_p(r,t)=\sum_{n=0}^\infty{R_n\over r^n},
\ee
where $n=1$ term is equivalent to 
our perturbation for $\gamma\approx 1$.
Had we used this perturbation instead of (\ref{r}) we would have
obtained the standard growing modes 
at the third order of the 
perturbation ({\it i.e.,} at $n=3$) \cite{elcio}.
Therefore, our analysis shows that, 
in spite of the growing modes,       
the perturbation cannot grow to 
scales much larger than the
correlation length, in a flat universe. 
On the contrary, we shall see that in an open
universe a growing mode exists 
for these scales.

\begin{figure}[htb]
\begin{center}
\leavevmode
\epsfxsize=5truecm\rotatebox{-90}{\epsfbox{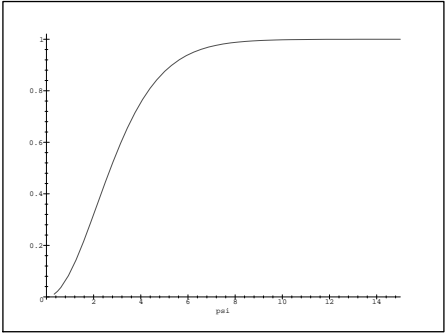}}
\end{center}
\caption{The graph of standard density contrast 
($\gamma=0$) verses $\psi$ for an open universe. 
\label{fig:densitycontrast-standard}}
\end{figure}

\begin{figure}[htb]
\begin{center}
\leavevmode
\epsfxsize=5truecm\rotatebox{-90}{\epsfbox{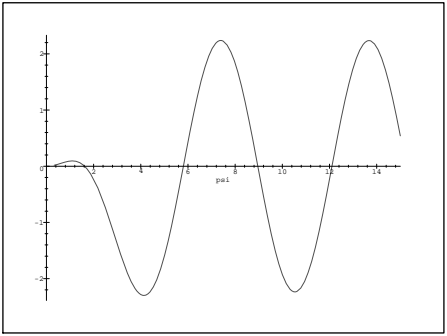}}
\end{center}
\caption{The graph of density contrast, at 
$\gamma=1$ verses $\psi$, for an open universe . 
\label{fig:densitycontrast-ours}}
\end{figure}

The differential equation (\ref{yequation}) for
a non-flat universe, $k\not = 0$, is more complicated. 
To start with, we consider a matter-dominated universe 
where both the homogeneous background 
pressure and the perturbative contributions 
from the inhomogeneities vanish 
({\it i.e.}\  $\alpha=0\, ,\, \omega=0$). 
The Friedmann scale factor is
given parametrically by
\be
R={ {\rm d}t\over {\rm d}\psi}\quad;\quad
t={\Omega\over 2H(1-\Omega)^{3/2}} 
(\psi-{\rm sin}\psi),\label{rparameter}
\ee
where $\psi$ is real for an open universe and 
imaginary, $\psi=-i\theta$, for a closed 
universe, and $\Omega$ is the ratio of the present to the
critical density \cite{weinberg,kolb,lucchin}. The differential equation 
(\ref{yequation}) is now a 
hypergeometric equation which can be solved
exactly. The density contrast is given by the growing mode
\bear
\delta_+(\psi)&\sim&\frac{\sin\left(\sqrt\gamma\psi\right)}
{\sqrt\gamma}\left[
\frac{-3\sinh\psi}{\left(\cosh\psi-1\right)^2}
+\frac {2\gamma\sinh\psi}
{\cosh\psi-1}\right] \nonumber\\
&+&\cos\left(\sqrt\gamma\psi\right)
\frac{5+\cosh\psi}{\cosh\psi-1} \quad ,
\label{delta-plus-open} %%(16)
\eear
and the decaying mode
\bear
\delta_-(\psi)&\sim&\cos\left(\sqrt\gamma\psi \right)\left[
\frac{-3\sinh\psi}{\left(\cosh\psi-1\right)^2} +\frac {2\gamma\sinh\psi}
{\cosh\psi-1}\right] \nonumber\\
&-&\sqrt\gamma\sin\left(\sqrt\gamma\psi\right)
\frac{5+\cosh\psi}{\cosh\psi-1}\quad . 
\label{delta-minus-open} %%(17)
\eear
We note that for $\gamma=0$,
the above modes coincide with
the usual 
results in the literature
\cite{weinberg,lucchin}. 
Contrary to the 
decay of fluctuations in the flat universe, 
scale-invariant inhomogeneities can grow 
in an open matter-dominated universe. 
Infact, unlike the standard open universe
density contrast which remains constant at a redshift of 
approximately $2/5\Omega$ ({\it i.e.}, ${\rm cosh}\psi\le 5$, 
Fig. \ref{fig:densitycontrast-standard}), our density 
contrast oscillates with $\psi$  
(Fig. \ref{fig:densitycontrast-ours}).


\begin{figure}[htb]
\begin{center}
\leavevmode
\epsfxsize=5truecm\rotatebox{-90}{\epsfbox{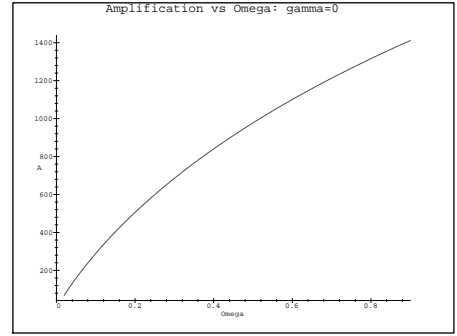}}
\end{center}
\caption{The graph of standard amplification factor 
($\gamma=0$) verses $\Omega$. 
\label{fig:afactor-standard}}
\end{figure}

For the amplification factor, $\delta(t_0)/\delta(t_r)$, 
from the time of recombination to now, once again we have the standard 
results for $\gamma=0$ in which the amplification 
rises monotonically with $\Omega$ (Fig. \ref{fig:afactor-standard}).
\begin{figure}[htb]
\begin{center}
\leavevmode
\epsfxsize=5truecm\rotatebox{-90}{\epsfbox{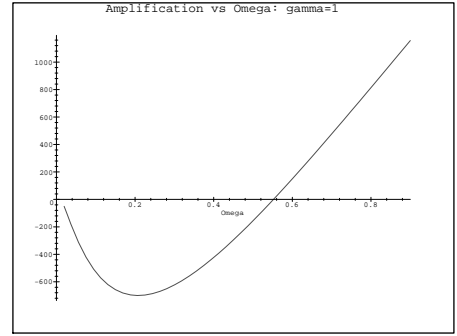}}
\end{center}
\caption{The graphs of amplification factor 
at $\gamma=1$ verses $\Omega$ 
\label{fig:our amplification factor}}
\end{figure}
\begin{figure}[htb]
\begin{center}
\leavevmode
\epsfxsize=5truecm\rotatebox{-90}{\epsfbox{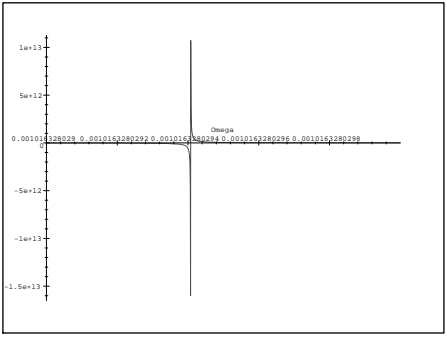}}
\end{center}
\caption{The graphs of amplification factor 
at $\gamma=1$ for small values of $\Omega$. 
\label{fig:our}}
\end{figure}
However, for $\gamma=1$,  we see that the amplification
varies with $\Omega$ in an oscillatory manner 
(Fig. \ref{fig:our amplification factor}). Furthermore,
going to a very high resolution, at small values of Omega, we see
a distinct peak at $\Omega\approx 0.001$ where an amplification
by up to a factor of $10^{13}$ is obtained (Fig. \ref{fig:our}). 
The amplification is finely tuned to
the value of $\Omega$. 
For example an amplification of $10^5$ occurs for 
$\Omega\approx 0.00101629$. At higher values of $\Omega$ no
peaks and at much lower values
of $\Omega$ many more peaks are observed.
This remarkable phenomenon does not occur
in the standard analysis at $\gamma=0$. Indeed, a
self-similar or hierarchical universe is expected to
have a very low density which is why
our perturbation decays in a flat universe 
\cite{pietroneroreport}. 
Here, we have shown that for
the growth of a scale-invariant inhomogeneity 
to be in accordance with
the anisotropy of the microwave background radiation, 
an upper limit of $\Omega\approx 0.001$ is 
required for the density of a hierarchical universe. 

\begin{figure}[htb]
\begin{center}
\leavevmode
\epsfxsize=5truecm\rotatebox{-90}{\epsfbox{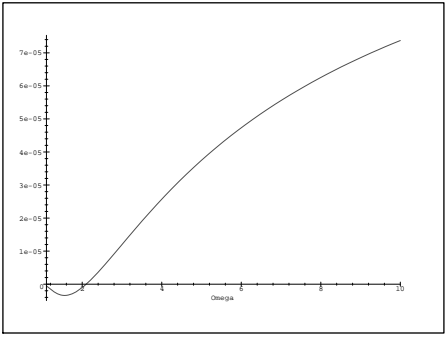}}
\end{center}
\caption{The graphs of amplification factor 
at $\gamma=1$ for a closed universe, $\Omega >1$. 
\label{fig:close}}
\end{figure}

The above results can be
verified further by studying the density contrast and amplification
factor in a closed universe. In the standard analysis,
the amplification factor increases with increasing $\Omega$ 
\cite{weinberg,kolb,lucchin}.
Using equations
(\ref{delta-plus-open}) and (\ref{delta-minus-open}) 
for a closed universe ({\it i.e.} $\psi=-i\theta$), we see that
the decaying mode (\ref{delta-minus-open}) is 
imaginary for $\Omega >1$ and the 
growing mode (\ref{delta-plus-open}) 
now decays. The graph of amplification
factor shows that the perturbation decays significantly 
from the recombination to 
the present time (Fig. \ref{fig:close}). It also shows that
the decay factor increases with increasing $\Omega$. 

For a matter-dominated universe with a 
non-vanishing perturbation pressure ({\it i.e.}\ 
$\omega=0$ but $\alpha\not =0$)  
the differential equation (\ref{yequation}) is no longer a
hypergeometric equation. However, it can be
transformed into a hypergeometric equation
by using appropriate transformations \cite{morse-feshbach}.
The solution is given by 
the hypergeometric function,
\bear
Y& =&\left({\rm sin}{\theta\over 2}\right)^{-3\alpha}\nonumber\\
&\times &F\!\left(-{3\alpha+1\over 2}+a,
\,- {3\alpha+1\over 2}-a ,\,
-{6\alpha+1\over 2}\, ;\, {\rm sin}^2\frac\theta 2\right),\nonumber\\
& &
\eear
where $a=\sqrt{1-4A_3}/2$ and $\theta=-i\psi$. 
The second solution is obtained 
by replacing $\alpha$ with $-1-\alpha$ in the above.
Asymptotically, we obtain
\bear
\delta R&\sim& t^x\qquad ;\qquad x
=-\frac 32\alpha +\frac 12\sqrt{1-4A_3},\label{17}\\
\delta &\sim&t^{-\frac 32\alpha
+\frac 12\sqrt{1-4A_3}},
\eear
For complex values of the square-root term
in the density contrast, above, there is no 
growing mode since $\alpha$ is always positive. For real 
values of the square root, which 
occur for  $\alpha\mathop{>}\limits_\sim 0.8$ or 
$\alpha\mathop{<}\limits_\sim -0.4$ with $\gamma=1$, the situation is 
more complex. The terms $1-4A_3$ can now be 
written as $9\alpha^2-4\alpha-4$ whose square root is always smaller 
than $3\alpha$. We, therefore, conclude that asymptotically 
inhomogeneities cannot grow in an open
universe with a vanishing background 
and a non-vanishing
perturbation pressure.

Unlike the preceding cases, 
an exact solution cannot be obtained for
a universe with non-vanishing pressure. In this case, 
the Friedmann scale factor 
is given by the expression
\be
t=- {2 C_1R^{\frac 32(1+\omega)}\over 3(\omega+1)}
F\left(\frac 12 , {3(\omega+1)\over 2(3\omega+1)}, {9\omega+5\over
2(3\omega+1)};C_1^2 R^{1+3\omega}\right)
\ee
where $C_1$ is a constant. Although, the exact form of the above 
hypergeometric function is not unknown, it 
can be easily shown that it
reduces to 
(\ref{rparameter}) for a matter-dominated
universe and, asymptotically, to a linearly-increasing scale factor. 
In the asymptotic limit, the 
differential equation (\ref{yequation}) 
, for an open universe with $\omega > -1/3$, reads
\be
\frac{d^2Y}{dt^2} +Ct^{-3\omega-3}Y
+A_3t^{-2}Y =0 ,
\ee
where $C$ is a constant. 
This expression reduces to a confluent
hypergeometric equation for $F(1/\tau)$ after the redefinition :  
\be
Y=t^{\frac 12(1\pm\sqrt{1-4A_3})} e^{2\over (3\omega+1)\tau}F\left(
{1\over\tau}\right),
\ee  
where $t=C^{\frac 1 {1+3\omega}}\tau^{2\over 1+3\omega}$.
In the limit of large times, $F(1/\tau)$ 
approaches a constant value
\cite{morse-feshbach} and we recover expression (\ref{17})
whose behaviour does not depend on $\omega$. For the density
contrast, we obtain
\be
\delta\sim t^{-{3\alpha\over 2}+
3\omega+\frac 12\sqrt{1-4A_3}},
\ee
which shows that the perturbation grows for all positive values of
$3\omega-3\alpha/2$ or 
for $\omega\mathop{>}\limits_\sim -0.3$ and
positive values of $\alpha$.
The growth rate of the perturbation is significant in 
the radiation-dominated era ($t/t_0\approx 10^8$).
In this period, 
the perturbation amplifies by a
factor of $10^{4}$ for $\alpha\approx 1/3$. 
However, since the baryonic matter cannot grow in this era,
due to its coupling to the radiation, this result can 
only be relevant for non-baryonic dark matter.

The main results are summarized in table 1.
\vskip 0.2cm
\begin{tabular}{|c|c|c|}
\hline
& & \\
equation of state & $\qquad$ flat$\qquad$ & 
$\qquad$ open$\qquad$\\
$P=\omega\rho+\alpha\delta\rho r^{-\gamma}$ 
& $k=0$&$k=-1$\\
& &\\
\hline
& & \\
matter-dominated &{\it decays}  & {\it grows }\\
$\omega=0$ & &\\
& & \\
\hline
 & &\\
radiation-dominated &{\it decays} &{\it grows} \\
$\omega=\frac 13$ & &for large times  \\
& & \\
\hline
\end{tabular}
\vskip 0.2cm

Table 1:\,\,\,The table of density contrast 
for flat and open universes in 
matter and radiation-dominated eras.
\vskip 0.15cm

We have evaluated the density contrast for a
scale-invariant
perturbation in flat, open and closed universes. 
We have found that, 
in this specific perturbation scheme,
fluctuations decay in 
the Einstein-de-Sitter universe 
whereas they grow 
in an open universe.
The standard results for the growing and decaying
modes of the density contrast in open and
closed universes are contained in our results.
The decaying mode for a flat universe is also reproduced
by our method. However, the absence of the standard
growing mode in the flat universe shows that the perturbation 
grows to larger scales 
in an open universe than in a flat universe.
In addition, in our analysis, the perturbation 
does not simply amplify with increasing $\Omega$. 
It has an oscillatory behaviour. A remarkable
feature of our perturbation, which is not observed 
in the standard analysis,
is the existence of sharp peaks at $\Omega\le 0.001$.  
At $\Omega\approx 0.001$ the perturbation can be 
amplified by up to a factor of $10^{13}$, from 
the recombination to the present time. 
We further confirm our results by showing that
the perturbation decays even faster in a closed universe.
We, therefore, conclude that 
for scale-invariant density perturbations the standard 
belief that the density contrast grows faster 
with increasing $\Omega$ is not valid.
Our results are fully consistent with the 
low density expected for a hierarchical universe.

E.A thanks Conselho Federal de Desenvolvimento Cient\'\i fico e
Tecnol\'ogico (CNPq-Brazil) for financial support. R. M. was supported by
Funda\c c\~ao de Amparo a Pesquisa do
Estado de S\~ao Paulo (FAPESP) and The Abdus Salam International Center 
for Theoretical Physics (ICTP).
We thank M. B. Ribeiro for 
many helpful discussions and comments. 

%%%%%%%%%%%%%%%%%%%%%%%%%%%%%%%%%%%%%%%%%%%%%%%%%%%%%%%%%%%%%%%%%%%%%%%%%%%

\end{document}